\documentstyle[aps,prl,fancyheadings,epsf,multicol,amssymb]{revtex}

\def\bbbc{{\mathchoice {\setbox0=\hbox{$\displaystyle\rm C$}\hbox{\hbox 
to0pt{\kern0.4\wd0\vrule height0.9\ht0\hss}\box0}} 
{\setbox0=\hbox{$\textstyle\rm C$}\hbox{\hbox 
to0pt{\kern0.4\wd0\vrule height0.9\ht0\hss}\box0}} 
{\setbox0=\hbox{$\scriptstyle\rm C$}\hbox{\hbox 
to0pt{\kern0.4\wd0\vrule height0.9\ht0\hss}\box0}} 
{\setbox0=\hbox{$\scriptscriptstyle\rm C$}\hbox{\hbox 
to0pt{\kern0.4\wd0\vrule height0.9\ht0\hss}\box0}}}} 
\begin{document} 
 
\title{ 
STM measurement of single spin relaxation time in superconductors. 
} 
\author{Jurij \v{S}makov$^{1,2}$, Ivar Martin$^2$, 
and Alexander V. Balatsky$^2$} 
\address{$^1$Theoretical Physics, Department of Physics, Royal Institute of 
Technology, SE-10044 Stockholm, Sweden\\ 
$^2$Theoretical Division, Los Alamos National Laboratory, 
Los Alamos, NM 87545} 
 
\date{Printed \today} 
 
\maketitle 
 
\begin{abstract} 
Localized spin states in conventional superconductors at low temperatures  are 
expected to have long decoherence time due to the strong suppression of
spin relaxation channels. We propose a scanning tunneling microscopy (STM) 
experiment allowing the direct measurement of the decoherence time of a single 
spin in a conventional  superconductor.  The experimental setup can
be readily  applied to general-purpose spin-polarized STM and to local spin
relaxation  spectroscopy.  A possible extension of the setup to a quantum
information  processing scheme is discussed. 
\end{abstract} 
\begin{multicols}{2} 

When a magnetic atom is placed in a conventional $s$-wave  superconductor, the
exchange interaction between the localized spin and the  conduction electrons
leads to formation of {\em spin polarized} scattering resonances inside the
superconducting energy gap\cite{shiba,sbs,yazdani}.  These resonances manifest
themselves as sharp peaks in the density of states localized near  the
impurity.  The spin-relaxation time for such states is governed by the 
coupling to the environment.  At low temperatures, this coupling is strongly
suppressed due to the gapped nature of quasiparticles in $s$-wave
superconductors.  The weak spin-relaxation channels related to magnetic
dipole-dipole interaction with the nuclear magnetic moments can be further
reduced by choosing a superconductor with zero nuclear spin.  It is
therefore expected that long spin-relaxation times are achievable when a local
moment is placed in $s$-wave superconductor.  In this Letter we propose an
experimental setup capable of determining the single-spin relaxation time for
such defects.  Provided that the measured spin-relaxation time is sufficiently
long, superconducting tips ending with a single magnetic atom can be used for
low-temperature spin-polarized STM.  We also discuss a possible application of
the spin-polarized local states in superconductors in quantum information
processing schemes.

The tunneling current between two spin-polarized states depends strongly on 
the angle between the respective spin polarizations.  The time-dependence of
the tunneling current can hence be used to extract the dynamics of the
individual spins, assuming they are weakly coupled.  This is the basis for our
method of determination of the single-spin relaxation time in superconductors. 
In the proposed experiment, both sides of the tunnel junction are identical
superconductors, each containing one magnetic atom.  The spin-polarized signal
will be optimized when the impurities on the two sides of the junction are
spatially aligned; this is most easily achieved in the STM setup, when one of
the magnetic atoms is placed directly on the tip.  Superconducting (for
example, Nb) STM tips have already been successfully used\cite{davis} and the
STM capabilities for the single-atom manipulation \cite{gauthier} make the
fabrication of the tip ending with a single magnetic impurity atom feasible.  

To theoretically predict the dependence of the $I$-$V$ characteristic on the
angle between the impurity spins on the tip and on the surface, we use the
$T$-matrix approach, combined with the tunneling Hamiltonian formalism. 
We model the impurity by introducing a local perturbation to the  
Bardeen-Cooper-Schrieffer (BCS) Hamiltonian describing a homogeneous  
superconductor. In the Nambu formalism \cite{fetter}, the impurity Hamiltonian  
can be written in terms of two-component Nambu spinors, $\Psi_{\mathbf k}=
[c_{{\bf k}\uparrow}\ c^\dagger_{-{\bf k}\downarrow}]^T$,
as \cite{sbs} 
\begin{equation} 
H_{\textrm{imp}}=\frac{1}{N}\mathop{\sum}_{{\mathbf k}{\mathbf k}'} 
\Psi^{\dagger}_{\mathbf k}\hat{V}\Psi_{{\mathbf k'}}. 
\end{equation} 
Here $\hat{V}=W\hat{\tau}_0+U\hat{\tau}_3$, $\hat{\tau}_0$ is the $2\times 2$ 
unit matrix, $\hat{\tau}_i\, (i=1,2,3)$ are the Pauli matrices. The impurity 
strength is characterized by the exchange interaction with the conduction
electrons,  $W$, and the on-site potential, $U$, or the corresponding
dimensionless  parameters $w=\pi N_0W$ and $u=\pi N_0U$, where $N_0$ is the 
density of states at Fermi level in the normal state. 
We can then obtain the expression for the complete 
finite-temperature Green's function (GF) in the presence of single impurity,
\begin{eqnarray} 
\label{fullgf} 
\hat{\mathcal G}({\mathbf r},{\mathbf r'};i\omega)&=& 
\hat{\mathcal G}^{(0)}({\mathbf r}-{\mathbf r'};i\omega)+\nonumber\\ 
&+&\hat{\mathcal G}^{(0)}({\mathbf r};i\omega)\hat{T}(i\omega) 
\hat{\mathcal G}^{(0)}(-{\mathbf r'};i\omega),
\end{eqnarray}  
in terms of the GF for the homogeneous SC  
$\hat{\mathcal G}^{(0)}({\mathbf r};i\omega)$ and the $T$-matrix 
$\hat{T}(i\omega)=[\hat{V}^{-1}-\hat{\mathcal G}^{(0)}({\mathbf r}=0;i\omega)]^{-1}$. 
We approximate the finite-temperature   GFs for the homogeneous SC in the
real-space representation by 
\begin{equation} 
\label{gf0} 
\hat{\mathcal G}^{(0)}({\mathbf r};i\omega)= 
-\frac{\pi N_0[i\omega\hat{\tau}_0+ 
\Delta\hat{\tau}_1]}{\sqrt{\omega^2+\Delta^2}}\frac{\sin k_F r} 
{k_F r}, 
\end{equation}  
where $2\Delta$ is the SC energy gap and  
$k_F$ is the Fermi momentum. By using expression (\ref{gf0}) we neglect the  
exponential decay of the GF for the homogeneous superconductor with  
$r=|{\mathbf r}|$ on the length scale of the superconducting coherence  
length (typically $\sim 1000$ \AA\ for conventional SCs) and the effects of  
the finite bandwidth. Both approximations are justified, because we are only  
interested in short-length (of the order of few \AA ngstr\"oms) and  
low-frequency (of the order of SC gap) behavior of the GF. We also  
neglect the suppression of the SC order parameter in the vicinity of the 
impurity, because this approximation does not significantly affect the 
spectral functions \cite{sbs}. Using this set of approximations, we  
calculate the GF for a superconductor with an impurity and found the  
expressions for the spectral functions, corresponding to its diagonal  
elements $A_{ii}({\mathbf r},{\mathbf r'};\epsilon)=-(1/\pi)\textrm{Im}\, 
G^{R}_{ii}({\mathbf r},{\mathbf r'};\pm\epsilon)$. Here the plus  
sign in the argument of the retarded GF  
$G^{R}_{ii}({\mathbf r},{\mathbf r'};\epsilon)$, obtained by analytic  
continuation of (\ref{fullgf}), corresponds to $i=1$ and minus sign to $i=2$. 
Spectral functions $A_{ii}$ have  
a particle-hole-symmetric continuous part at frequencies $|\epsilon|>\Delta$  
and a $\delta$-function singularities on the symmetric energies,  
\begin{equation} 
\label{intra} 
\epsilon_0=(-)^i\frac{\alpha\Delta}{\sqrt{1+\alpha^2}}, 
\end{equation} 
where $\alpha=(1+u^2-w^2)/(2w)$. 
 
In the tunneling Hamiltonian formalism, the quasiparticle tunneling current can
be expressed  as a convolution of the spectral functions on different sides of
the junction  \cite{mahan}. In our case, the impurities on the tip and on the
surface  (we will use sub- or superscript ``$1$'' to refer to the tip  
properties and ``$2$'' for the surface properties) may  have different spin
orientations. To account for this, we introduce angle  $\theta$ between the
spin directions and express the fermion creation   and annihilation operators
on the surface in the spin basis with the quantization axis along
the tip impurity spin. The result is 
\begin{eqnarray} 
c^{\dagger}_{{\mathbf r}_2\uparrow}&=& 
\cos\frac{\theta}{2}c^{\dagger}_{{\mathbf r}_2\bar{\uparrow}}+ 
\sin\frac{\theta}{2}c^{\dagger}_{{\mathbf r}_2\bar{\downarrow}},\\ 
c^{\dagger}_{{\mathbf r}_2\downarrow}&=& 
-\sin\frac{\theta}{2}c^{\dagger}_{{\mathbf r}_2\bar{\uparrow}}+ 
\cos\frac{\theta}{2}c^{\dagger}_{{\mathbf r}_2\bar{\downarrow}}.\nonumber 
\end{eqnarray} 
Here $\uparrow$, for example, stands for the ``up'' spin orientation in the 
surface spin basis and $\bar{\uparrow}$ is  the ``up'' spin in the tip spin
basis. Then, we come to the expression   for the total quasiparticle current 
\begin{equation} 
\label{total} 
I(V)=(I_{11}+I_{22})\cos^2\frac{\theta}{2}+ 
(I_{12}+I_{21})\sin^2\frac{\theta}{2}, 
\end{equation} 
with the partial currents, $I_{ij}(V)$, given by 
\begin{eqnarray} 
\label{part} 
I_{ij}(V)&=& 
2\pi e\sum T_{{\mathbf r}_1,{\mathbf r}_2} 
T^{*}_{{\mathbf r}'_1,{\mathbf r}'_2}\int_{-\infty}^{+\infty} 
\Theta(\epsilon,eV)\times\nonumber\\ 
&\times&A^{(1)}_{ii}({\mathbf r}_1,{\mathbf r'}_1;\epsilon) 
A^{(2)}_{jj}({\mathbf r}_2,{\mathbf r'}_2;\epsilon-eV)\,d\epsilon. 
\end{eqnarray} 
Here $e$ is the electron charge, $\Theta(\epsilon,eV)= 
n_F(\epsilon-eV)-n_F(\epsilon)$ is the difference of the Fermi-Dirac  
distribution functions on the different sides of the junction,  $T_{{\mathbf
r}_1,{\mathbf r}_2}$ is the tunneling matrix element,  and the summation is
performed over   all lattice sites on the tip (${\mathbf r}_1$, ${\mathbf
r}'_1$) and on the   surface (${\mathbf r}_2$, ${\mathbf r}'_2$).   Expression
(\ref{part}) is for general electrode geometry. We can further  simplify it by
assuming that the tunneling happens from a single point  on the tip (the
impurity atom). The tunneling matrix element then takes  the form
$T(r)=T\exp(-\sqrt{r^2+z^2}/d)$ that depends on the lateral distance $r$ from
the projection of the tip onto the surface, tip-surface distance  $z$, and the
characteristic decay length $d$, related to the material work function
(typically about 0.5\AA).   Detailed calculations for this and other geometries
will be presented elsewhere. 

\begin{figure}[htbp] 
  \begin{center} 
  \epsfxsize=3.0in 
  \epsfbox{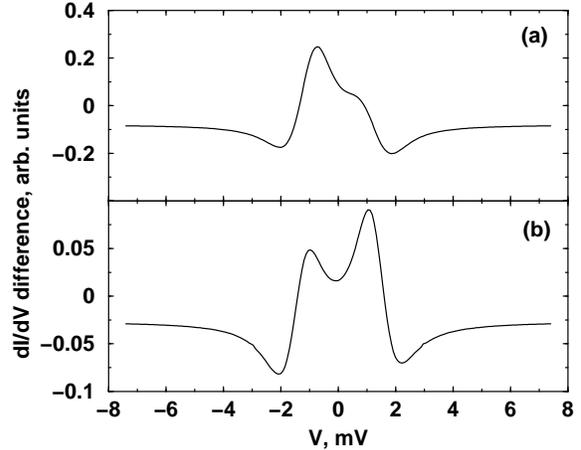} 
\caption{ 
The calculated change in the $dI/dV$ spectrum due to the presence of Mn (a) and
Gd (b) adatoms.  Impurity parameters were taken to be $w=0.9$, $u=1.0$ for Mn 
and $w=0.3$, $u=-0.3$ for Gd. The surface is Nb with $\Delta=1.48$ meV  at
$T=3.85$ K. Tip-surface distance is $z=5$ \AA\ and $d=0.5$ \AA.} 
\label{fig1} 
\end{center} 
\end{figure}

To calibrate the theoretical parameters for the impurity and the
superconducting host, we consider tunneling experiment from a $normal$ STM tip
into the surface of the SC (Nb) with the magnetic adatoms (Gd or Mn) on
it\cite{yazdani}.  We can qualitatively reproduce experimental tunneling
spectra by tuning the impurity parameters $w$ and $u$. Results of this
calculation are presented in   Fig. \ref{fig1}.  In what follows,
we adopt parameters characteristic of Mn adatom in Nb: $w=0.9$, $u=1.0$,
and $\Delta=1.48$ meV. 
 
\begin{figure}[bhtp] 
  \begin{center} 
  \epsfxsize=3.0in 
  \epsfbox{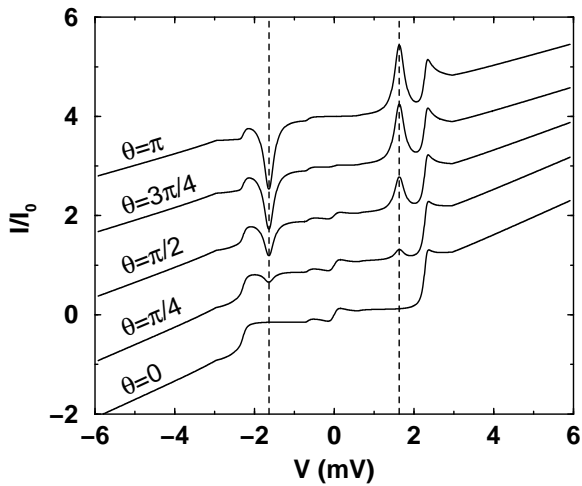} 
\caption{ 
Calculated $I$-$V$ characteristic of the Nb superconducting tunneling contact
for different values of angle $\theta$ between Mn spins on two sides of the
junction.  The current is normalized by $I_0=\sigma_0\Delta_1/e$,  where
$\sigma_0$ is the normal-state conductance of the tunneling junction.  The
curves are  offset by $I_0$ and the $\delta$-functions at  $V=\pm 1.63$ mV
(dashed vertical lines) are broadened for clarity. } 
\label{theta} 
\end{center} 
\end{figure}

When the impurities are present both on the SC tip and on the   SC surface, a
number of new features appear in the   $I$-$V$ characteristic. For example, 
when voltage first becomes large enough (in absolute value) for the   intragap
state of the tip to overlap with the continuous part of the spectrum
of the surface, the abrupt increase (or decrease) in current results. The
most remarkable feature of these $I$-$V$   characteristics, however, is their
strong dependence on the relative   orientation of the impurity spins, i.e. on
the angle $\theta$. The reason   for this strong dependence is the existence of
the contribution to the   current, resulting from the overlap between the
intragap states of the   tip and of the surface (at certain values of voltage).
From the   expression for the position of the intragap states,
Eq.~(\ref{intra}), it can be seen that these peaks (generally there are four of
them) appear at voltages 
\begin{equation} 
eV=\pm\frac{\alpha_1\Delta_1}{\sqrt{1+\alpha_1^2}}\pm 
\frac{\alpha_2\Delta_2}{\sqrt{1+\alpha_2^2}}. 
\end{equation} 
For the identical impurities in the identical superconductors,
($\alpha_1=\alpha_2$), $\Delta_1=\Delta_2$, at $\theta=0$ these peaks  vanish
because the corresponding intragap states, contributing to $I_{11}$ and
$I_{22}$ overlap only at zero voltage and hence there is no
tunneling current between them.  Fig. \ref{theta}  
presents the $I$-$V$ characteristics calculated for various values of
angle $\theta$.  The $\delta$-function peaks due to the tunneling between the
intragap states appear for non-zero $\theta$ at $V=\pm 1.63$ mV and become more
intensive with increasing $\theta$. It is this strong $\theta$ dependence that
allows us to formulate the experimental procedure for the measurement of the
decoherence time of the intragap impurity spin-polarized states in the SC. 
 
\begin{figure}[bhtp] 
  \begin{center} 
  \epsfxsize=3.0in 
  \epsfbox{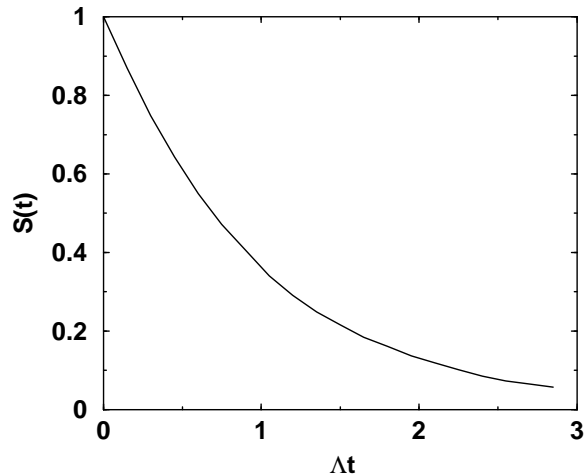} 
\caption{ Current-current correlation function  $S(t)=(\langle I(t)
I(0)\rangle-\langle I \rangle^2) /(\langle I^2\rangle-\langle I \rangle^2)$,
calculated for uncorrelated diffusion of spin orientations.  Assuming that
each spin diffuses over the unit sphere with the diffusion coefficient
$\Lambda$, the relative angle between the spins evolves according to stochastic
PDE, $d\theta = \Lambda \cot(\theta) dt/2 + \sqrt{\Lambda} d\xi$, where $\xi$
is the normal random variable with zero mean and variance $\langle
d\xi^2\rangle = dt$.  This correlation function is well approximated by the
usual expression for diffusion, $\exp(-\Lambda t)$. } 
\label{fig:St} 
\end{center} 
\end{figure}

The simplest way to measure the spin decoherence time is to monitor time
dependence of the tunneling current, $I(t)$, for applied voltage fixed at the
resonant value ($V =\pm 1.63$ mV in our calculation).  The current-current
correlation function, 
$S(t)=(\langle I(t) I(0)\rangle-\langle I \rangle^2)
/(\langle I^2\rangle-\langle I \rangle^2)$, can be used to directly extract the spin-relaxation time. 
The theoretical prediction for $S(t)$ is presented in Fig.~\ref{fig:St}.   
Another way to extract spin dynamics is to first polarize the spins by external
magnetic field (larger than $k_B T/\mu_B$ but smaller than $H_{c1}$) and then
observe the relaxation of the current after the field is removed.  The effect
of the magnetic dipole-dipole interaction between the impurity spins can be
studied by measuring the current relaxation at different tip-surface
distances.  It is possible, however, that the time resolution of the experiment
is insufficient to resolve the times associated with the spin dynamics; in this
case, the measured current will be a constant equal to the average value in
Eq.~(\ref{total}).  

Generally, the same technique may be used to measure the relaxation time of 
\emph{any} local spin state on a conducting substrate, as long as this
relaxation time is shorter than the relaxation time of the impurity-induced
intragap state in the SC tip, e.g. Kondo spin dynamics on a metallic
surface\cite{mano}. 
 
\begin{figure}[bhtp] 
  \begin{center} 
  \epsfxsize=3.0in 
  \epsfbox{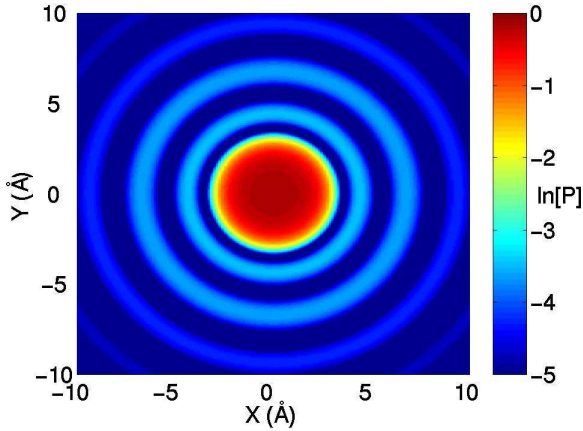} 
\end{center} 
\caption{ 
The topographic map of the log of the spin-contrast STM signal $P$ in the  
vicinity of the magnetic impurity.  Rapidly decaying Friedel oscillations are
clearly seen at large distances.  For this calculation we have taken the same
values of parameters as for Fig.~\ref{theta} and fixed the bias voltage at
$V=1.63$ mV.} 
\label{contrast} 
\end{figure}

If the relaxation time of the intragap spin-polarized impurity state in  a SC
proves to be sufficiently long, a SC tip ending with a magnetic impurity can be
used to obtain the spin-contrast STM images.  Currently, to achieve the
spin contrast in the STM measurements one has to use ferromagnetic tips
\cite{wiesen,kirschner}.  These STM experiments have already
produced a number of interesting results, like the direct imaging of the
two-dimensional antiferromagnetism \cite{wiesen}.  Unfortunately, in such
setup, the increase of the tunneling area of the tip due to the
restrictions on the tip material leads to the degradation of the spatial
resolution, compared to conventional STM.  In addition, strong magnetic fields
produced by the tips may affect the measurements by inducing polarization of
the local spin structure, or in the case of superconducting samples, cause a
local SC-to-normal transition.  Our proposed setup would be the next step in
the spin-polarized STM, which would allow the spin-contrast imaging at low
biases with an atomically sharp STM tip.  In this setup, the effect of the
magnetic field due to the magnetization of the tip is minimized since a single
impurity spin would be sufficient to achieve the spin contrast. As an example,
in Fig. \ref{contrast}, we present a theoretically calculated map of the
spin-contrast signal,
\begin{equation} 
\label{p} 
P=\frac{I_{\uparrow\uparrow}-I_{\uparrow\downarrow}} 
{I_{\uparrow\uparrow}+I_{\uparrow\downarrow}},
\end{equation} 
in the vicinity of surface impurity. Currents in Eq.~(\ref{p}) correspond to the
parallel ($I_{\uparrow\uparrow}$) and antiparallel ($I_{\uparrow\downarrow}$) 
orientations of the impurity spins on the tip and on the surface.  

Another application of the spin polarized states near magnetic atoms in  
superconductors is related to quantum information processing.  It is motivated
by the expected long relaxation times of spins in $s$-wave superconductors and
the ease of read-out of the spin states using the spin-polarized tunneling
techniques discussed above.  At temperatures much lower than the SC transition
temperature, the quasiparticle contribution to the spin relaxation will
effectively disappear.  The dipolar relaxation due to interaction with the
lattice nuclear spins can be also significantly reduced by using SC with
non-magnetic nuclei, e.g. non-magnetic isotopes of Pb or Sn.  A
local spin in the $s$-wave superconductor is isolated from environment to the
same extent as a local spin in a semiconductor with the band gap equal to
$2\Delta$.  An advantage compared to current proposals based of semiconductors,
e.g. phosphorus in silicon\cite{kane}, is that the substrate is conductive and
hence one can do local tunneling experiments to interrogate local spins that
store and process quantum information.  The details of the proposed quantum
information procession architecture based on the local spins in superconductors
will be discussed elsewhere\cite{ivar}. 
  
In summary, we have proposed an STM experiment, which will allow the  
measurement of the decoherence times of the single spins in superconductors.  
For sufficiently long measured spin-decoherence time, the
setup  can be applied to the spin-resolved STM measurements.  Also, the
possibly long decoherence time of the single-impurity spin states in
superconductors, combined with the ease of readout and manipulation, make such
states a favorable candidate for a qubit in quantum computing applications. 

We thank Don Eigler and Gerard Milburn for stimulating discussions. J. \v{S}.
would like to thank the Los Alamos National Laboratory for hospitality and
gratefully acknowledge partial financial support from the Swedish Research 
Council.

\end{multicols} 
\end{document}